\documentstyle[12pt,aaspp4]{article}
\begin{document}
\parindent=1.0cm

\title{Deep $g'r'i'z'$ GMOS Imaging of the Dwarf Irregular Galaxy \\ Kar 50
\footnote[1]{Based on observations obtained at the
Gemini Observatory, which is operated by the Association of Universities
for Research in Astronomy, Inc., under a co-operative agreement with the
NSF on behalf of the Gemini partnership: the National Science Foundation
(United States), the Particle Physics and Astronomy Research Council
(United Kingdom), the National Research Council of Canada (Canada),
CONICYT (Chile), the Australian Research Council (Australia), CNPq (Brazil),
and CONICET (Argentina).}}

\author{T. J. Davidge}

\affil{Canadian Gemini Office, Herzberg Institute of Astrophysics,
\\National Research Council of Canada, 5071 West Saanich Road, 
\\Victoria, B.C. Canada V9E 2E7\\ {\it email: tim.davidge@nrc.ca}}

\begin{abstract}

	Images obtained with the Gemini Multi-Object Spectrograph (GMOS) 
are used to investigate the stellar content and distance 
of the dwarf irregular galaxy Kar 50. The brightest object is an HII region, 
and the bright stellar content is dominated by stars with $g'-r' < 0$. The 
tips of the main sequence and the red giant branch are tentatively identified 
near $r' = 24.9$ and $i' \sim 25.5$, respectively. The galaxy has a blue 
integrated color with no significant color gradient, and we conclude that Kar 
50 has experienced a recent galaxy-wide episode of star formation. The 
distance estimated from the brightest blue 
stars indicates that Kar 50 is behind the M81 group, 
and this is consistent with the tentative RGB-tip brightness.
Kar 50 has a remarkably flat central surface brightness profile, even at 
wavelengths approaching $1\mu$m, although there is no evidence of a bar. 
In the absence of another large star-forming 
episode, Kar 50 will evolve into a very low surface brightness galaxy.

\end{abstract}

\keywords{galaxies: individual (Kar 50) - galaxies: dwarf - galaxies: stellar content} 

\section{INTRODUCTION}

	The Local Group contains over 30 dwarf galaxies that span a range of 
morphological types (e.g. van den Bergh 2000). While such nearby systems are 
fundamental laboratories for investigating dwarf galaxy evolution, 
many are satellites of M31 and the Milky-Way, and 
hence have been affected by hierarchal interactions 
(e.g. Mayer et al. 2001). In order to develop a more 
comprehensive understanding of dwarf galaxies, including the effects 
of environment, it is necessary to study objects outside of the Local Group.

	Located at a distance of $\sim 3.5$ Mpc, the M81 group 
contains a number of dwarf galaxies, some of which are systematically 
different from those in the Local Group (Caldwell et al. 1998). 
Karachentseva, Karachentsev \& B\"{o}rngen (1985, hereafter KKR) list possible 
dwarf members of the M81 group. Many of these have yet to be studied in 
detail, and some will likely turn out not 
to be members of the M81 group. This being said, the census of galaxies in the 
M81 group is likely far from complete, and Froebrich \& Meusinger (2000) report 
the detection of additional possible low surface brightness members.

	One of the faintest, and hence intrinsically most 
interesting, systems in the KKR compilation is Kar 50. KKR noted that Kar 
50 is `unusually filamentary with barely outlined knots', and they 
partially resolved Kar 50 into stars using images 
recorded during 1 arcsec seeing conditions. 
Various observational properties of Kar 50 are summarized in Table 1.

	Motivated by the possible intrinsic faintness of Kar 50, coupled with 
its unusual appearance, we decided to include this galaxy in a background field 
obtained as part of a program to study the outer regions of NGC 2403. The 
observations and the data reduction procedures are discussed in \S 2, and these 
data are used to investigate the integrated photometric properties of 
Kar 50 in \S 3. Our $i'$ and $z'$ images are of particular interest 
for work of this nature, as they sample redder wavelengths 
than previous studies, and hence are better probes of any 
cool stellar component, such as might arise from an old population. The 
properties of the brightest stars in Kar 50 are 
investigated in \S 4. A distance based on the brightest stars is calculated 
in \S 5, and we conclude that Kar 50 is located well behind the M81 group. 
A brief summary and discussion of the results follows in \S 6.

\section{OBSERVATIONS \& REDUCTIONS}

	The data were recorded with the Gemini Multi-Object Spectrograph (GMOS) 
on the Gemini North telescope during 2 nights in November 2001 as part of a 
system verification program to investigate the outer regions of NGC 2403 
(Davidge \& Roy 2002, in preparation). Readers are refered to 
Crampton et al. (2000) for a detailed description of GMOS. 
The GMOS detector consists of three buttable $2048 \times 4608$ EEV 
CCDs, with each pixel subtending 0.0727 arcsec on a side. The sky field 
has a 5.5 arcmin diameter, although the detector covers a much 
larger area so that spectra of objects near the edge of the sky field can be 
obtained.

	Details of the observations, including dates, 
exposure times, and image quality are listed in Table 2. The sky conditions 
were photometric when the data were obtained, and standard stars were observed 
on each night. The CCDs were binned $2 \times 2$ during readout to better 
match the seeing conditions, so that each superpixel covers 0.145 arcsec on a 
side. Bias frames and twilight flats were also obtained on each night. 

	The data were reduced using tasks in the 
Gemini IRAF package. The reduction pipeline included (1) removal of 
CCD-to-CCD gain differences, (2) subtraction of the bias 
pattern, (3) division by a twilight sky flat, and (4) mosaicking 
of the processed CCD output to form a single image. 
Interference fringes were removed from the $i'$ and $z'$ data 
by subtracting a fringe frame that was created by combining observations 
of different fields. The processed images were then aligned and median-combined 
on a filter-by-filter basis to reject cosmic rays.

\section{THE APPEARANCE AND INTEGRATED PHOTOMETRIC PROPERTIES OF KAR 50}

	The portion of the final $r'$ image containing Kar 50 is shown 
in Figure 1. These data have a finer angular resolution 
and go deeper than previous observations of Kar 50. There is an absence of 
structures such as a bar or a central nucleus. 
Concentrations of stars are evident in the outer envelope, and 
these are likely the `filamentary' features refered to by KKR. There 
are a number of stars evident throughout the galaxy, 
although the density of sources outside of the main body of 
the galaxy is significant, and it can be anticipated that many of the 
objects detected in Kar 50 may be foreground stars and background galaxies.

	Karachentseva et al. (1987; hereafter K87) included Kar 50 in their 
investigation of the structural properties of M81 group dwarf galaxies. Kar 
50 is the most compact object in their sample, and falls within the 
scatter envelope of the relations between size, surface brightness, and total 
intrinsic brightness defined by low surface brightness dwarf spheroidals. 
Both K87 and Barazza, Binggeli, \& Prugniel (2001, hereafter BBP) find that the 
central surface brightness profile of Kar 50 at visible and red wavelengths is 
very flat, changing by only a few tenths of a magnitude per square arcsec 
within a few arcsec of the galaxy center. The surface 
brightness profile of the halo follows an exponential relation (BBP). 

	A series of aperture measurements, positioned on the geometric 
center of Kar 50 as established from the $i'$ image, were made in each filter. 
A technique similar to that employed by Davidge (1992) and Piotto, King, 
\& Djorgovski (1988) to investigate the color profiles of 
globular clusters was used to suppress bright resolved stars. In particular, 
succesive annuli were divided into six azimuthal segments, and the median 
surface brightness within each group of segments was assigned to that annulus. 

	The resulting $g',r',i'$ and $z'$ surface brightness profiles are shown 
in Figure 2. The error bars show the uncertainty in the surface brightness 
measurements in the last annulus due to uncertainties in the local sky level, 
which were estimated by examining the background at various locations in 
the vicinity of Kar 50. The light profiles are remarkably flat near the galaxy 
center, with a scatter of only $\pm 0.04$ mag in each filter in the central 8 
arcsec. Light profiles obtained without azimuthal median filtering are 
similar to those in Figure 2. 

	The light profiles measured in $i'$ and $z'$ are of particular 
interest, as previous studies of Kar 50 have not sampled wavelengths 
longward of $\sim 0.7\mu$m. Not only are the $i'$ and $z'$ data 
less susceptible to reddening than the $g'$ 
and $r'$ observations, but they also are a more sensitive probe of 
any underlying old and/or intermediate-age population. It is evident from 
Figure 2 that there are no hints of a central concentration in the main body 
of the galaxy, even in $i'$ and $z'$.

	The $g'-i'$ and $r-z'$ color profiles of Kar 50 
are shown in Figure 3. The color profiles do not show significant 
gradients when $r < 15$ arcsec, and the only deviations from a flat 
color profile occur at larger radii, where the uncertainties in the color 
measurements become significant. The onset of the exponential light 
profile near 8 arcsec is not accompanied by an obvious change in stellar 
content. The stellar content of Kar 50 is thus well mixed throughout 
the galaxy.

	Lenz et al. (1998) investigate the photometric properties of stars in 
the SDSS filter system, and colors predicted from their log(g) = 4.5 solar 
metallicity sequences for three effective tempertures are marked in Figure 3. 
There is a tendency for the observations at longer wavelengths 
to sample cooler populations, as expected for a composite stellar system. 
Nevertheless, the relatively blue $(r'-z')$ colors indicate 
that the integrated light from Kar 50 is dominated by hot young stars, 
even near $1\mu$m. 

\section{THE RESOLVED STELLAR CONTENT}

	The current investigation is restricted to objects within $\sim 30$ 
arcsec of the center of Kar 50, and photometry of sources over the remainder 
of the GMOS field of view will be discussed in a future paper dealing 
with the outer regions of NGC 2403 (Davidge \& Roy 2002, in preparation). 
The brightnesses of individual stars were measured with the 
PSF-fitting routine ALLSTAR (Stetson \& Harris 1988), using co-ordinates, 
initial brightnesses, and PSFs obtained from DAOPHOT (Stetson 1987) tasks. The 
PSFs were constructed from moderately bright stars located over 
the entire GMOS field of view. 

	The $(r', g'-r')$ and $(i', r'-i')$ CMDs of stars in two radial 
intervals centered on Kar 50 are shown in Figures 4 and 5; we do not consider 
the $z'$ measurements here as they do not go as deep as the other 
data. The CMDs labelled `Kar 50' contain sources within 22 arcsec 
of the galaxy center, while the CMDs labelled `background' consist of 
sources between 22 to 31 arcsec from the galaxy center; the former region 
has the same area as the latter. The radial boundaries used to 
construct the CMDs are marked in Figure 1.

	The number density of the brightest stars in the left hand panels of 
Figure 4 and 5 is only slightly higher than that of background objects having 
comparable brightness. At fainter levels the contrast with 
respect to background sources increases, and there is a concentration of stars 
in the Kar 50 CMDs with relatively blue colors $(g'-r' \sim 
-0.4$, and $r'-i' \sim 0$) near $r' \sim i' \sim 25$. Based on the blue colors, 
we identify this clump of stars as the onset of the upper main sequence. The 
$r'$ and $i'$ observations probe a redder population than the $g'$ data, 
and there is a clump of stars near the faint end of the $(i', r'-i')$ CMD 
of Kar 50 at $i' \sim 25.5$ and $r'-i' \sim 0.3$. This color is consistent with 
these objects being red giants, and we tentatively identify this collection 
of stars as the RGB-tip. The identifications of the main sequence and RGB are 
discussed further in \S 6.

	Contamination from foreground stars and background galaxies complicates 
efforts to study stars brighter than $r' \sim i' = 25$ in Figures 4 and 5, 
but color information can be used to identify 
a sample of bright stars that likely belong to Kar 50. Stars occupy a distinct 
region of the $(g'-r', r'-i')$ two-color diagram (TCD) (e.g. 
Krisciunas, Margon, \& Szkody 1998), and multi-color observations can be 
used to identify non-stellar interlopers (e.g. Fan 1999). The $(g'-r', r'-i')$ 
TCDs of the Kar 50 and background fields are compared in Figure 6; the 
majority of objects plotted in this figure have $r' < 25.5$. 
The solid line shows the predicted locus of solar-metallicity log(g) = 
4.5 stars from Table 4 of Lenz et al. (1998), while 
the dashed line is the ridgeline of the SDSS stellar sequence from Figure 3 of 
Schneider et al. (2001); the data points in Figure 6 
agree better with the SDSS sequence than the model sequence.

	The objects within 22 arcsec of Kar 50 plotted in Figure 6 tend to be 
bluer than those in the background, in the sense that the majority of 
sources in Kar 50 have $(g'-r') < 0$, whereas only two objects in the 
background region are this blue. Given this obvious difference, it was 
assumed that all sources with $g'-r' < 0$ within 22 arcsec of Kar 50 belong 
to the galaxy, while the sources with $g'-r' > 0$ are likely contaminating 
objects. This criterion for Kar 50 membership is further verified by comparing 
the number of objects with $g'-r' > 0$ in the right and left hand panels of 
Figure 6: there are 14 sources with $(g'-r') > 0$ within 22 arcsec of the 
galaxy center, and 15 such objects in the background field. 
This excellent agreement indicates that Kar 50 likely does not harbour a 
substantial component with $r' < 25$ and $(g'-r') > 0$; 
rather, hot stars appear to dominate the bright stellar content of Kar 50.

	The $(r', g'-r')$ and $(i', r'-i')$ CMDs of Kar 50, corrected for 
line of sight absorption using $E(B-V)$ determined from the Schlegel, 
Finkbeiner, \& Davis (1998) maps, are shown in Figure 7. The imposition of a 
selection criterion based on $g'-r'$ has a major impact on the interpretation 
of the CMDs when $r < 25$, in that many of the sources that might have been 
selected as candidate bright members of Kar 50 in the absence of $g'-r'$ 
color information likely do not belong to the 
galaxy; the impact is most dramatic on the $(i', r'-i')$ 
CMD, where sources with $g'-r' > 0$ are indicated. The hot stars in Kar 
50 with $r_0 \sim 24.8$ that we associate with the upper main sequence have 
$(g'-r')_0 \leq -0.4$. The solar metallicity models of Lenz et al. (1998) 
predict that the blue limit for stars is $g'-r' \sim -0.5$, and 
so it appears that upper main sequence stars in Kar 50 are not subject to 
internal reddening in excess of E$(g'-r') = 0.1$ mag. 

	The brightest blue object in the $(r', g'-r')$ CMD of Kar 50, the 
location of which is indicated in Figure 1, has $r' = 22.7$ and $g'-r' = -0.3$. 
While the $g'-r'$ color is within the acceptable range for stellar sources,
this object has an extremely blue $r'-i$ color, with $r'-i' 
\sim -0.9$, placing it in a region of the $(g'-r', r'-i')$ TCD that is 
populated by compact emission line galaxies (Fan 1999). 
This object is likely an HII region in Kar 50, with the $g'$ and $r'$ 
measurements dominated by strong [OIII] and H$\alpha$ emission, respectively. 
This object is not extended in the 0.6 arcsec FWHM $r'$ 
and $i'$ images, indicating that it has a linear size $< 40$ parsecs 
using the distance derived in the next Section. As the only large 
HII region in Kar 50, this object has the potential of providing unique 
information about the gas-phase chemical composition of the galaxy. 

\section{THE DISTANCE TO KAR 50}

	The integrated colors and stellar content of Kar 50 indicate 
that the galaxy has experienced recent star formation, and so 
the properties of the brightest blue stars provide a means of estimating 
the distance to this system. Rozanski \& Rowan-Robinson (1994) discuss the 
brightest stars as distance indicators, and give a relation between the 
mean magnitude of the three brightest blue stars and galaxy brightness. The 
calibration in their Figure 10c has $\pm 1$ mag scatter, largely 
because the bright blue stellar content (1) is subject to stochastic effects, 
(2) is sensitive to the recent star-forming history, and (3) may be 
affected by internal extinction. Despite the large scatter in the calibration, 
the brightest blue stars still provide interesting limits on 
the distance of Kar 50, which are consistent with other indicators (\S 6).

	The brightnesses and colors of the three brightest objects 
with stellar spectral-energy distributions (SEDs) in Kar 50 
are summarized in Table 3. The $V_0$ and 
$(B-V)_0$ entries for each star are in the AB$_{95}$ system, and these were 
computed from the transformation equations given in Section 2.1 of Krisciunas 
et al. (1998). We prefer the Krisciunas et al (1998) transformation 
relations over those given by Fukugita et al. (1996) as the former 
are based on actual stellar observations; nevertheless, a potential 
source of uncertainty is that the brightest 
stars in Kar 50 are supergiants, whereas the transformation relations are 
likely dominated by main sequence and giant branch stars. Adopting B$_{AB95} = 
-0.12$ for Vega (Fukugita et al. 1996) then B$_0$(3) = 23.3 in the Johnson 
system. 

	The relation between M$_{B}$(3) and M$_{B_{T}}$ in Figure 10c of 
Rozanski \& Rowan-Robinson (1994) was iterated until convergence, 
and the Kar 50 distance modulus was found to be 30.4; 
the integrated brightness of Kar 50, computed from 
the B$_T$ value in Table 1, is then M$_B = -12.7$. The uncertainty in these 
quantities is dominated by the scatter in the calibrating relation, and so we 
assign $\pm 1$ mag errors to the distance modulus and M$_B$.

\section{DISCUSSION \& SUMMARY}

	Deep multicolor images obtained with the GMOS on the Gemini North 
telescope have been used to investigate the structure and stellar content 
of the dwarf irregular galaxy Kar 50. The presence of a significant population 
of bright blue stars, coupled with the blue integrated colors of the galaxy 
and the flat color profiles, which indicate that 
the stellar content of the galaxy is well mixed, 
indicate that Kar 50 has experienced a recent galaxy-wide star-forming episode. 
Kar 50 may thus be one of the small percentage of dwarf irregular galaxies 
that at any given time host a strong burst of star formation (Dohm-Palmer et 
al. 1998). In the remainder of this Section we discuss (1) the 
stellar content of Kar 50, with emphasis on checking the distance 
estimated from the brightest blue stars, and (2) the evolution of the galaxy. 

\subsection{The Distance and Stellar Content of Kar 50}

	While contamination from foreground stars and background galaxies 
complicates efforts to study the resolved stellar content of Kar 50 when 
$r' < 25$, in \S 4 it is demonstrated that a secure population of bright galaxy 
members can be identified using $g'-r'$ color. The application 
of a color criterion to reject contaminating objects will of course 
cause some objects, in this case those that have red colors, 
to be excluded from the analysis. The brightest red supergiants (RSG)s 
in dwarf galaxies have M$_V \sim -7$ (Rozanski \& Rowan-Robinson 1994), and 
thus would be expected to occur near $r' \sim 23.7$ at the distance of Kar 50. 
While this is well within the detection limits of our data, the 
agreement between the numbers of objects with $g'-r' < 0$ in the Kar 50 and 
background fields indicates that RSGs do not occur in great numbers in Kar 50. 

	The brightest object in Kar 50 has an SED suggesting that it is 
an HII region, and if this is the case then the $r'$ flux will be dominated 
by H$\alpha$ emission. With an assumed distance modulus of 30.4, this object 
has a luminosity of $1.3 \times 10^{37}$ ergs/sec in $r'$, which is consistent 
with the peak H$\alpha$ luminosity of HII regions in other irregular galaxies 
(Youngblood \& Hunter 1999). With a distance modulus of 27.5 then the 
luminosity of the HII region drops to $8.9 \times 10^{35}$ erg/sec, which is 
near the lower end of what is seen in irregular galaxies. Thus, the luminosity 
of this object is consistent with it being an HII region. Nevertheless, this 
object is very compact, as the maximum size of 40 parsecs is a factor of 
2 smaller than what is seen among HII regions in other irregular 
galaxies (Youngblood \& Hunter 1999).

	The distance of $12_{-4}^{+7}$ Mpc measured from the brightest 
blue stars places Kar 50 well behind the M81 group. There are significant 
uncertainties inherent to the use of the brightest blue stars as standard 
candles, and so in the remainder of this section we check if this distance is 
consistent with the properties of other resolved objects in Kar 50. 

	The RGB-tip is a prime distance indicator for old stellar systems, as 
it produces a clear break in luminosity functions and CMDs. The 
RGB-tip occurs near M$_{i'} \sim -4$ in old metal-poor populations (e.g. 
Davidge et al. 2002), and hence would occur at $i' \sim 23.5$ if Kar 50 were in 
the M81 group. Given the evidence for moderately recent star formation, a well 
populated AGB sequence might also be anticipated above the RGB-tip in Kar 50. 
There is no evidence in the CMDs of Kar 50 of either a well-defined RGB peaking 
near $i' \sim 23.5$ or a statistically significant AGB sequence when $i' < 
23.5$, which would produce an excess red population with respect to the 
background at this brightness. Moreover, it is evident 
from Figure 7 that many of the sources on 
the $(i', r'-i')$ CMD have $(r'-i') \sim 0.0$, whereas RGB stars will have 
$r'-i' \geq 0.3$. There is an apparent onset of stars in the 
$(i', r'-i')$ CMD with $i' \sim 25.4$ and $r'-i' \sim 0.3$. If this is 
the RGB-tip then this supports a distance modulus $\sim 29.4$, which falls 
within the $\pm 1$ mag errors in the distance modulus computed from the 
brightest blue stars. We are reluctant to assign a firm distance estimate based 
on the RGB-tip because it is at the faint limit of our data, and AGB stars 
can complicate distances measured from this feature.

	The brightest stars at visual wavelengths in star-forming galaxies 
tend to have spectral types A -- F (e.g. Humphreys \& McElroy 1984), and 
the stars used to estimate the distance to Kar 50 have colors that are 
consistent with this range of spectral types (\S 5). 
Humphreys \& McElroy (1984) list the effective 
temperatures and brightnesses of supergiants and upper main sequence stars in 
the Milky-Way and Magellanic Clouds, and we have used these data to locate 
these sequences on the $(r', g'-r')$ CMD. Colors were 
calculated using the relation between effective temperature and $g'-r'$ 
predicted by the solar metallicity log(g)=4.5 atmosphere models of Lenz et 
al. (1998), while M$_{r'}$ was computed from the Krisciunas et 
al. (1998) transformation relations. The Lenz et al. (1998) models have 
surface gravities appropriate for main sequence stars, but not supergiants, 
and this likely introduces systematic errors in $g'-r'$ on the order of a few 
hundredths of a magnitude in the placement of the supergiant sequence. 

	The loci defined by Ia and Ib supergiants, as well as the main 
sequence, are shown in Figure 8 based on the adopted distance modulus of 30.4. 
The type Ia supergiant sequence loosely matches the upper envelope of stars 
in Kar 50, while the Ib sequence passes through the main body of the CMD. 
In addition, the bright end of the main sequence is in reasonable agreement 
with the clump of stars we identified in \S 4 as the upper main sequence tip. 
Based on these comparisons we predict that spectroscopic 
studies will reveal that 1) the brightest stars in Kar 50 are 
Ia supergiants, and (2) the brightest main sequence stars in Kar 50 are 
early-type O stars.

\subsection{Some Comments on the Structure and Evolution of Kar 50}

	The light and color profiles of Kar 50 offer clues about the past 
evolution of the galaxy. The light profile of Kar 50 is flat over a 
linear scale of $\sim 1$ kpc, while the color profile suggests that 
the stellar content is well mixed throughout the galaxy. Light and 
color profiles similar to those in Kar 50
are not uncommon among Magellanic Irregular galaxies in the Virgo 
(e.g. Figures 8 and 9 of Binggeli \& Cameron 1993) and M81 (Bremnes, Binggeli, 
\& Prugniel 1998; Froebrich \& Meusinger 2000) groups. 

	Bar instabilities can mix gas throughout 
a system (e.g. Friedli, Benz, \& Kennicutt 1994), 
and produce flat light profiles. The disruption of bars may play an important 
role in the morphological evolution of dwarf galaxies, with the final system 
having a lower density and different kinematical properties (Mayer et al. 2001).
However, despite the flat light and color profiles of Kar 50, there is no 
evidence of a bar in Figure 1.

	Most surveys of the structural characteristics of nearby dwarf systems 
have relied on observations at blue and visible wavelengths. The $z'$ 
filter samples a wavelength interval that is less affected by recent 
star formation than filters covering shorter wavelengths, although the 
relatively blue $i'-z'$ color of Kar 50 suggests that a significant fraction of 
the light near $1\mu$m still comes from a young component. $JHK$ observations 
of Kar 50 will be of interest, as they will allow firmer constraints to be 
placed on (1) the size of any old stellar subtrate, (2) the degree of central
concentration of this component, and (3) the presence of a bar, given that 
bars are more easily detected in $H$ than at optical wavelengths 
(Eskridge et al. 2000).

	In the absence of subsequent star-forming episodes, 
Kar 50 will evolve into a very low surface brightness system. 
To estimate the surface brightness of Kar 50 after fading 
we adopt, based on the integrated colors of the galaxy, a M/L = 0.5 solar, 
and further assume that the final M/L = 1.0, based on the 
integrated color of a `typical' low surface brightness system in the Virgo 
cluster, for which $\overline{B-V} = 0.6$ (e.g. Impey, Bothun, \& Malin 1988). 
Kar 50 will then fade by 0.8 mag per square arcsec if the population ages 
passively, and the central surface brightness will drop to $\sim 25$ mag 
arcsec$^{-2}$ in $g'$. This is within the range of central surface brightnesses 
seen among low surface brightness galaxies in Virgo, some of which also 
have flat central light profiles (e.g. Impey et al. 1988).

\acknowledgements

Sincere thanks are extended to Jean-Rene Roy and Sidney van den Bergh for 
discussions regarding bars and the morphological properties of dwarf irregular 
galaxies.

\clearpage
\begin{center}
FIGURE CAPTIONS
\end{center}

\figcaption
[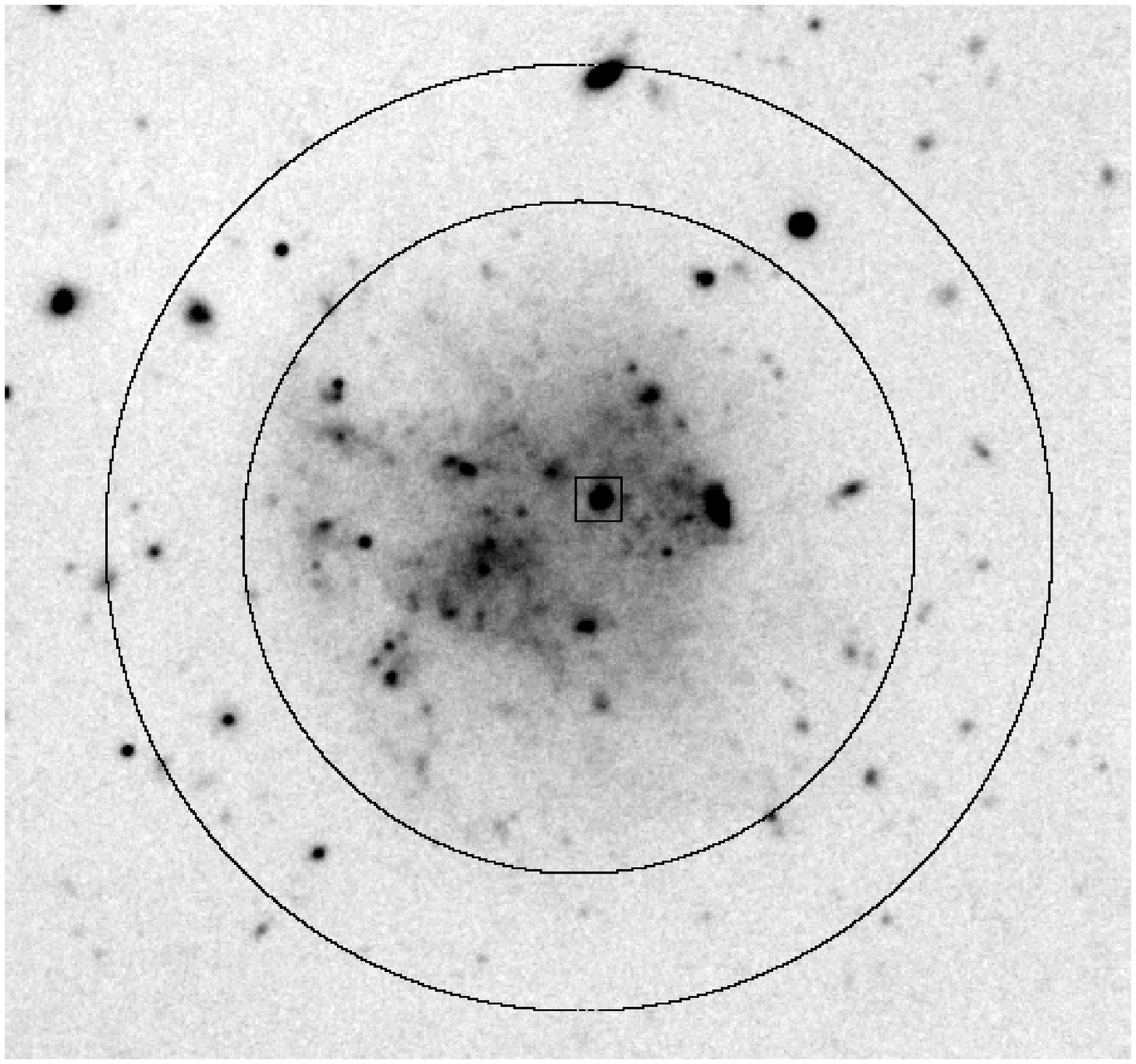]
{The portion of the final $r'$ image containing Kar 50. 
North is to the top and East is to the left. The square 
marks the HII region discussed in \S 4, while the two circles 
mark the intervals used in the stellar content investigation, which 
have outer radii of 22 and 31 arcsec.}

\figcaption
[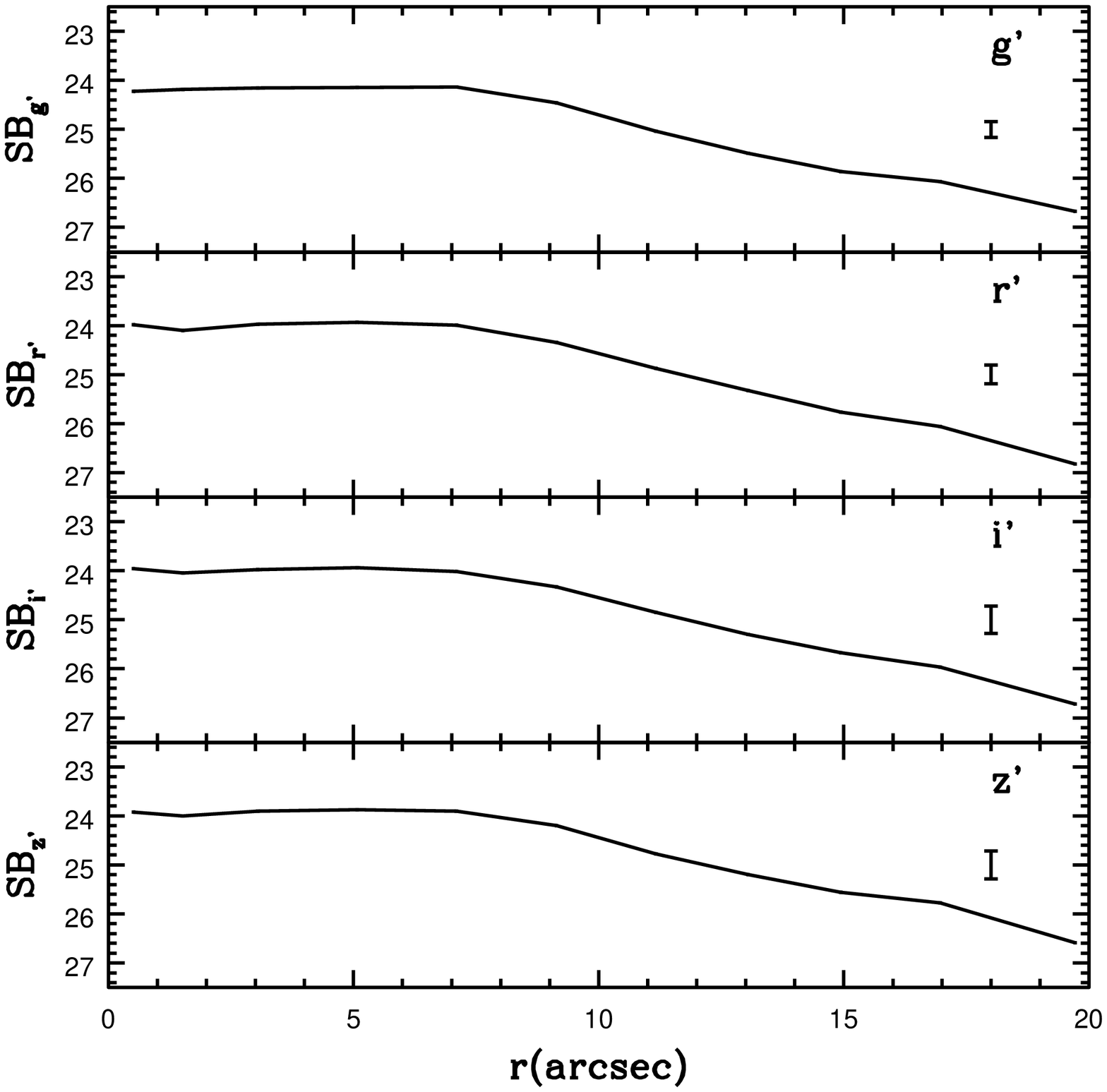]
{The $g', r', i'$, and $z'$ surface brightness profiles of Kar 50, 
obtained using the azimuthal filtering technique described in the 
text to suppress bright resolved stars. Note the remarkably flat profiles 
within 8 arcsec of the galaxy center. The error bars show the uncertainty in 
the largest radii measurements due to uncertainties in the sky level.} 

\figcaption
[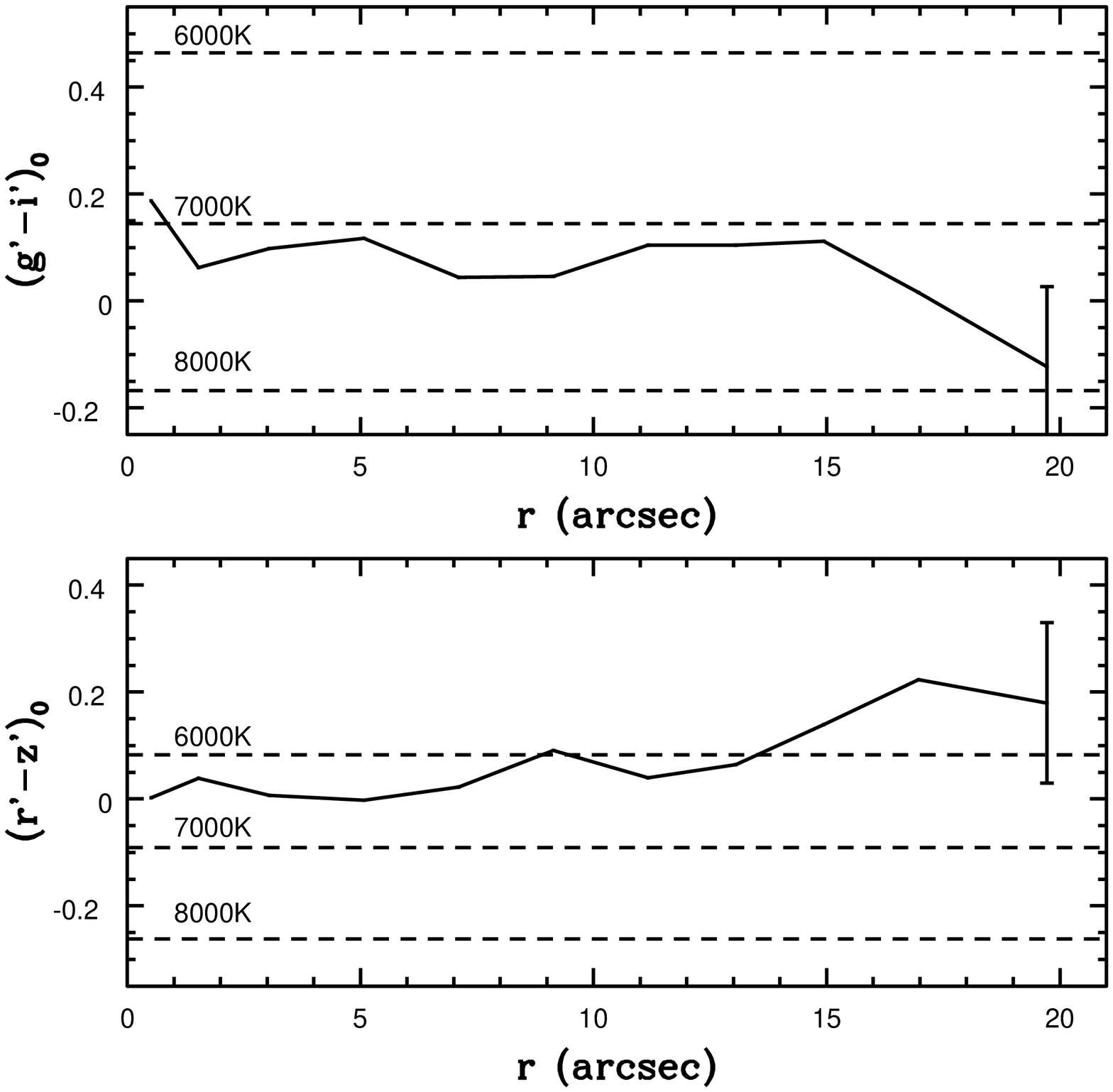]
{The $(g'-i')$ and $(r'-z')$ color profiles of Kar 50, corrected for line of 
sight reddening. The error bars show the uncertainty in the largest radii 
measurements due to uncertainties in the sky level, and the errors 
in the colors become progressively smaller than this with decreasing radius. 
The dashed lines show colors predicted for various effective temperatures from 
the log(g) = 4.5 solar metallicity models of Lenz et al. (1998). Note that 
there is no evidence for significant gradients within 15 arcsec of the galaxy 
center. While apparent departures from a flat color profile occur at larger 
radii, these are not significant given the uncertainty in the colors at these 
radii.}

\figcaption
[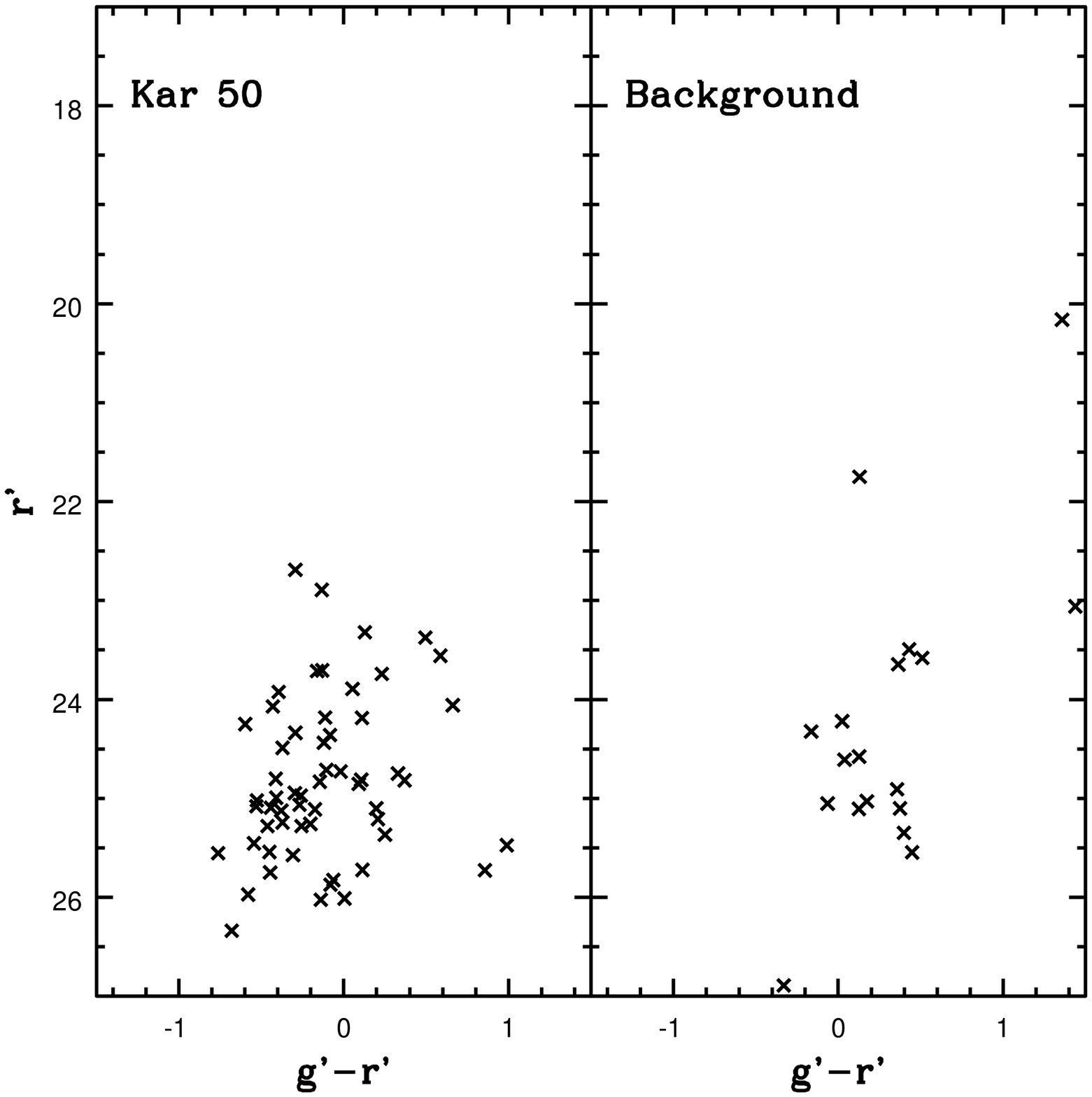]
{The $(r', g'-r')$ CMDs of objects within 22 arcsec of the center 
of Kar 50 (left hand panel), and those located between 22 and 31 arcsec from 
the galaxy center (right hand panel).}

\figcaption
[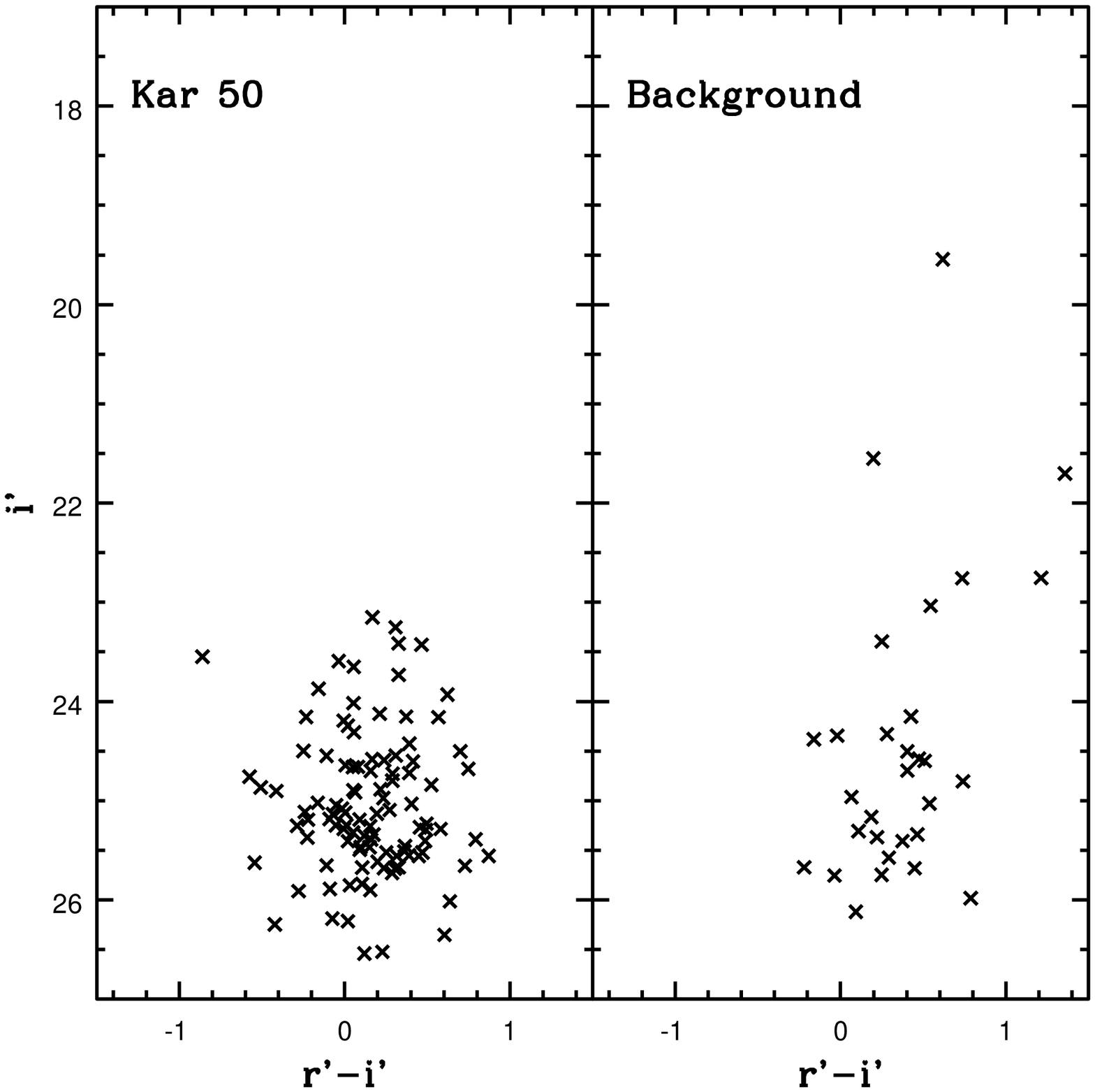]
{The $(i', r'-i')$ CMDs of objects within 22 arcsec of the center 
of Kar 50 (left hand panel), and those located between 22 and 31 arcsec 
from the galaxy center (right hand panel).}

\figcaption
[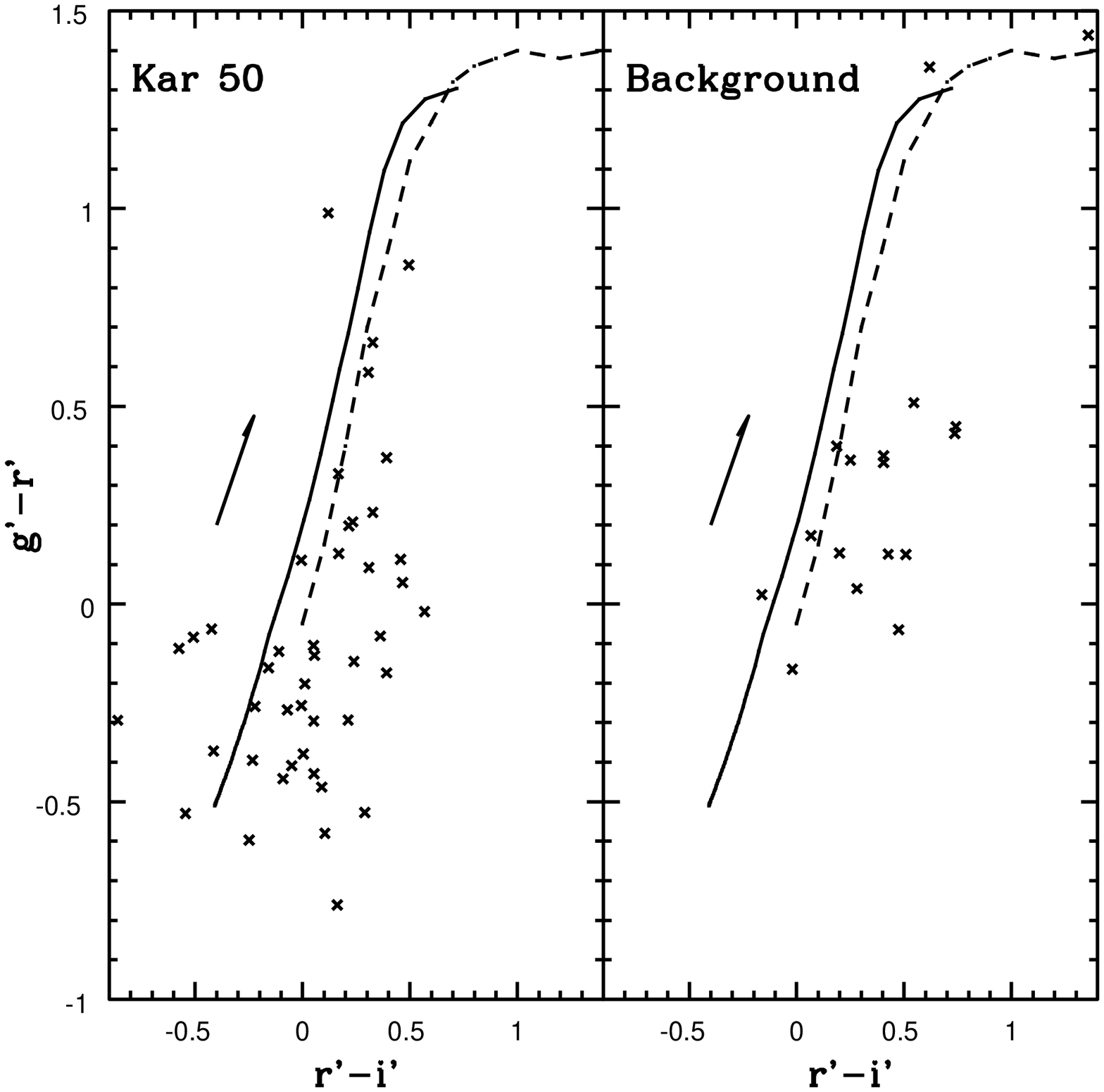]
{The $(g'-r', r'-i')$ TCDs of objects within 22 arcsec of the center 
of Kar 50 (left hand panel), and those located between 22 and 31 arcsec 
from the galaxy center (right hand panel). The solid line shows the locus of 
the solar metallicity log(g)=4.5 models of Lenz et al. 
(1998), while the dashed line is the SDSS stellar sequence from Figure 3 of 
Schneider et al. (2001). The reddening vector follows the relations 
listed in Table 6 of Schlegel et al. (1998), which in turn is based on the 
Cardelli, Clayton, \& Mathis (1989) and O'Donnell (1994) reddening curves. The 
length of the reddening vector corresponds to 
A$_{g'} = 1$ mag. Note that the majority of sources within 22 arcsec of the 
center of Kar 50 have $g'-r' < 0$, whereas those in the background field have 
$g'-r' > 0$.}

\figcaption
[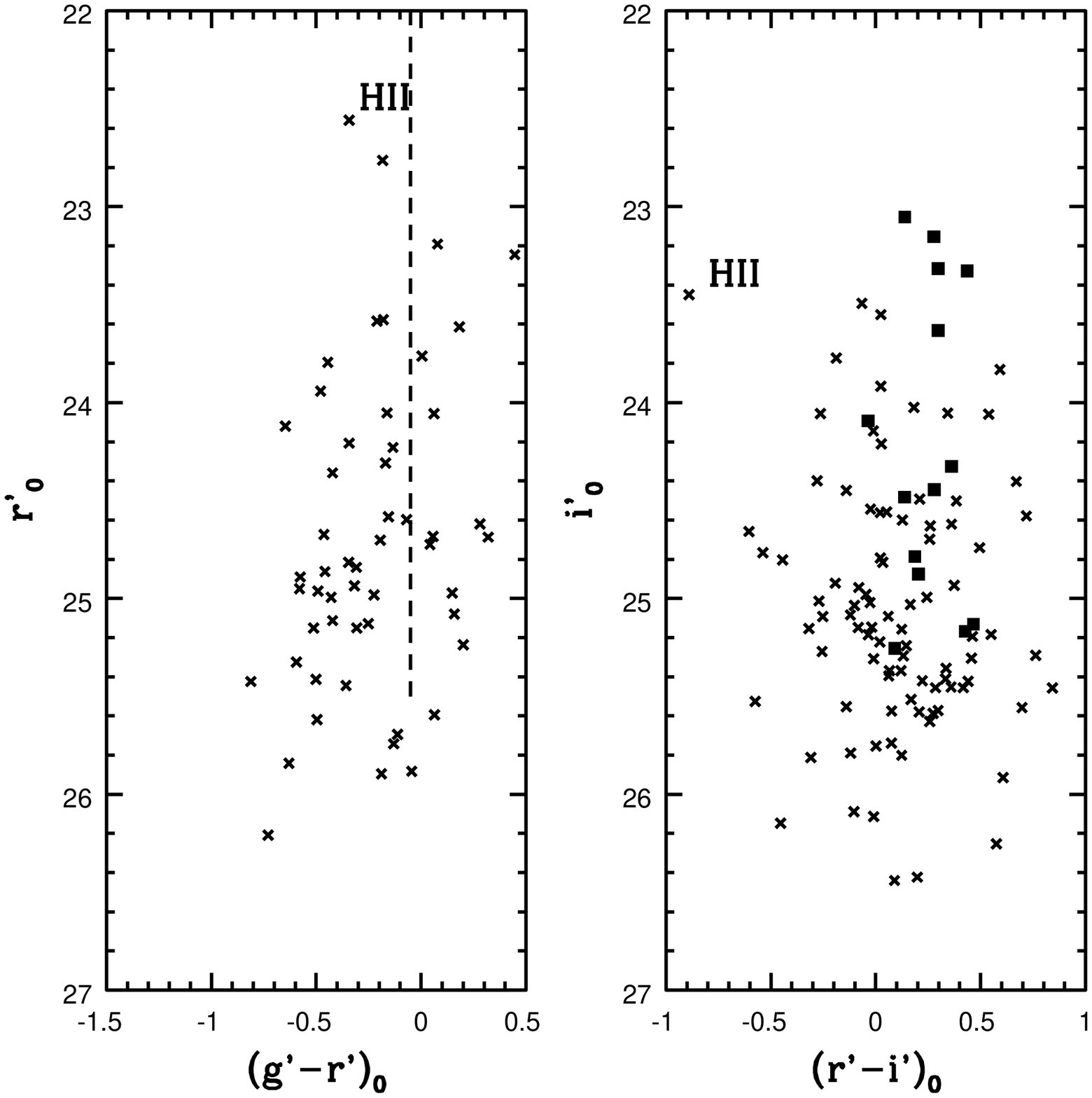]
{The $(r', g'-r')$ and $(i', r'-i')$ CMDs of stars within 22 arcsec 
of the center of Kar 50. The data have been corrected for line of sight 
extinction predicted from the Schlegel et al. (1998) maps. The dashed line 
marks $g'-r' = 0$, which is the criterion adopted for rejecting background 
objects; sources falling to the right of this line in the $(r', g'-r')$ CMD 
have a high likelihood of being contaminating objects. The solid 
squares in the right hand panel mark those sources 
detected in $g', r',$ and $i'$ having $g'-r' > 0$, and as such are probable 
foreground stars or background galaxies. The point corresponding to the 
candidate HII region is also indicated in each CMD.} 

\figcaption
[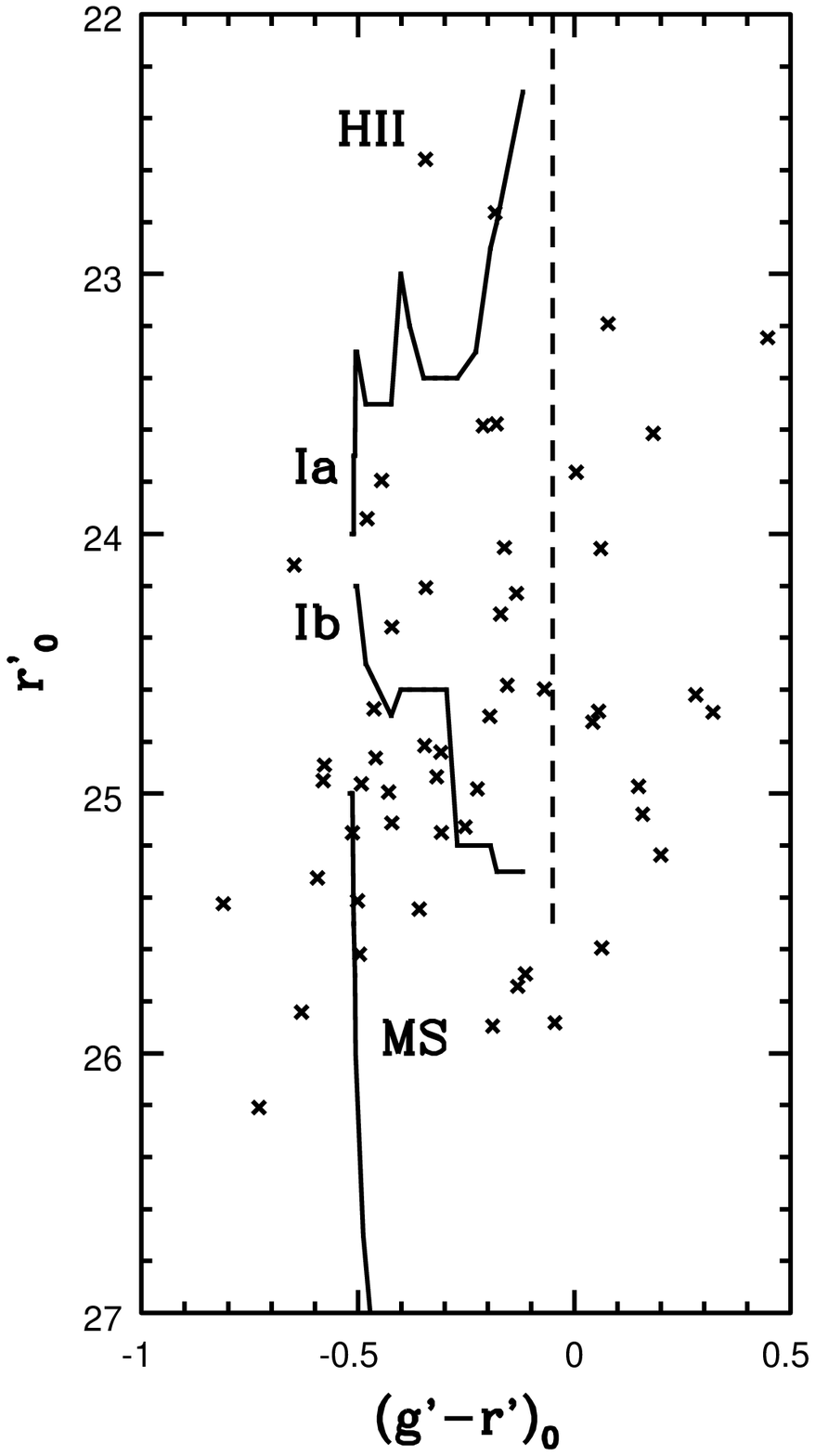]
{The $(r', g'-r')$ CMD of stars within 22 arcsec 
of the center of Kar 50. The data have been corrected for line of sight 
extinction as predicted from the Schlegel et al. (1998) maps. The dashed line 
marks $g'-r' = 0$; sources falling to the right of this line 
are likely unrelated to Kar 50. The source identified as an HII 
region is labelled, as are the loci of supergiants 
and upper main sequence stars in the Milky-Way and the 
Magellanic Clouds, assuming a distance modulus of 30.4.}
 
\clearpage

\begin{table*}
\begin{center}
\begin{tabular}{lcl}
\tableline\tableline
Quantity & Value & Reference \\
\tableline
RA (1950) & $07^{h}48^{m}25^{s}$ & KKR \\
Dec (1950) & $61^{o}31^{'}08^{"}$ & KKR \\
$l_{II}$ (1950) & 155.4789 & \\
$b_{II}$ (1950) & 30.8665 & \\
Type & dImV & KKR \\
B$_T$ & 17.88 & BBP \\
$r^{B}_{eff}$ & 11.1 arcsec & BBP \\
B--V & 0.24 & BBP \\
E(B--V) & 0.05 & Schlegel et al. (1998) \\
\tableline
\end{tabular}
\end{center}
\caption{Basic properties of Kar 50}
\end{table*}

\clearpage

\begin{table*}
\begin{center}
\begin{tabular}{cccc}
\tableline\tableline
Filter & Date & Exposure time & FWHM \\
 & (UT) & (sec) & (arcsec) \\
\tableline
$g'$ & Nov 22, 2001 & $2 \times 900$ & 1.00 \\
$r'$ & Nov 19, 2001 & $6 \times 400$ & 0.65 \\
$i'$ & Nov 19, 2001 & $6 \times 400$ & 0.65 \\
$z'$ & Nov 22, 2001 & $12 \times 200$ & 0.85 \\
\tableline
\end{tabular}
\end{center}
\caption{Details of Observations}
\end{table*}

\clearpage

\begin{table*}
\begin{center}
\begin{tabular}{ccccc}
\tableline\tableline
$r'_0$ & $(g'-r')_0$ & $(B-V)_0^a$ & $V_0^a$ & $B_0$ \\
\tableline
22.76 & --0.19 & --0.01 & 22.66 & 22.65 \\
23.57 & --0.18 & 0.00 & 23.47 & 23.46 \\
23.58 & --0.21 & --0.03 & 23.46 & 23.43 \\
\tableline
\end{tabular}
\caption{Photometric Properties of the Three Brightest Stars}
\end{center}
\tablenotetext{a}{Computed from $r'_0$ and $(g'-r')_0$ using the relations 
given in Section 2.1 of Krisciunas et al. (1998).}
\end{table*}

\end{document}